# Direct Exoplanet Investigation using Interstellar Space Probes[1]

Ian A. Crawford
Department of Earth and Planetary Sciences, Birkbeck College, University of London, UK.
E-mail: i.crawford@bbk.ac.uk

**Abstract**

Experience in exploring our own solar system has shown that direct investigation of planetary bodies using space probes invariably yields scientific knowledge not otherwise obtainable. In the case of exoplanets, such direct investigation may be required to confirm inferences made by astronomical observations, especially with regard to planetary interiors, surface processes, geological evolution, and possible biology. This will necessitate transporting sophisticated scientific instruments across interstellar space, and some proposed methods for achieving this with flight-times measured in decades are reviewed. It is concluded that, with the possible exception of very lightweight (and thus scientifically limited) probes accelerated to velocities of ~$0.1c$ with powerful Earth-based lasers, achieving such a capability may have to wait until the development of a space-based civilization capable of leveraging the material and energy resources of the solar system.

**Introduction**

The discovery that planets are ubiquitous companions of stars (as outlined in other chapters in this Handbook) naturally prompts consideration of how we might learn more about them. Such considerations become especially pressing when we realize that, although thousands of individual exoplanets have now been detected, our knowledge of them is generally limited to quite basic observational properties. These include orbital parameters (i.e. orbital semi-major axes, periods, and eccentricities), masses and/or radii (and thus bulk densities if both the latter values have been determined), and, in a handful of cases, some knowledge of atmospheric compositions.

There is an interesting parallel here with the exploration of our own solar system. Prior to the space age the basic physical properties of the planets had largely been determined by telescopic observations from the Earth. Information thus obtained included the orbital parameters, planetary radii, existence of moons (which, if present, allowed planetary masses to be determined), and, where appropriate, atmospheric compositions. However, although such observations were key steps in the exploration of the solar system, planetary science underwent a revolution once it became possible to make *in situ* observations using space probes. As a result, planetary data have become available that simply could not have been obtained using telescopes from the Earth. Given that this is true for the solar system, it seems clear that the same logic must hold for the study of exoplanetary systems as well.

---

[1] This is the submitted version of an invited review chapter now accepted for publication in *The Handbook of Exoplanets*, edited by H. J. Deeg and J. A. Belmonte, to be published by Springer International Publishing. The final version, possibly with minor changes, will be available from Springer (http://www.springer.com/gb/book/9783319553320) in due course.

It is of course true that, notwithstanding the wide range of potential scientific benefits, achieving interstellar spaceflight on timescales relevant to human society will be a formidable technological and societal undertaking. As even the closest exoplanets are several light-years away, it will be necessary for spacecraft to attain velocities that are a significant fraction of the speed of light if we are to visit them with travel times of decades. This is not a capability that exists at present and, as discussed below, may not exist for decades or centuries to come. Nevertheless, while acknowledging that such possibilities are at the most forward-looking end of the spectrum of exoplanet investigation techniques, we here review both the scientific case for direct exoplanet investigation using space probes and the range of suggested technological implementations.

## The scientific case for interstellar space probes

As discussed by Webb (1978), Wolczek (1982) and Crawford (2009), the overall scientific case for considering the construction of interstellar space probes can naturally be divided into four main areas:

- Studies of the interstellar medium (ISM), together with other astronomical investigations able to use an interstellar vehicle as an observing platform.

- Close-up astrophysical studies of the target star (or stars) that go beyond what can realistically be performed by solar system-based telescopes.

- *In situ* geological and planetary science investigations of exoplanetary systems.

- *In situ* biological studies of any indigenous lifeforms that may be present.

Although of considerable scientific interest, the topics covered by the first two bullet points lie beyond the scope of this Handbook. Here, we concentrate on the specifically planetary science and astrobiological benefits that may be expected should interstellar spaceflight prove feasible.

*Planetary science*

The diversity of planetary bodies in our solar system, awe-inspiring though it is, seems likely to pale in comparison with what awaits us in other planetary systems. Already, astronomical observations have demonstrated considerable exoplanet diversity, with entire classes of planets (e.g. 'hot Jupiters', 'super Earths', and planets tidally-locked to their host stars) that are not represented in the solar system at all. Moreover, where mass and radius data are available, it has become clear that many Earth-sized exoplanets are not even approximately Earth-like, with estimated compositions spreading the entire gamut from mostly iron to mostly water (Rappaport et al 2013, Dressing et al 2015), and still more exotic possibilities doubtless exist (e.g. Madhusudhan et al 2012). Moreover, all this diversity in bulk planetary properties is likely to be just be the tip of the iceberg: the diversity in climates, geological processes, and surface morphologies can, at present, only be imagined.

Investigating, and ultimately understanding, this diversity is of key interest to planetary scientists, and may eventually enable the development, and even the empirical testing, of general theories of planetary evolution that is not possible with the small sample of planets in the solar system. Although a start to cataloguing exoplanet diversity can be made by astronomical observations, and these capabilities will certainly improve greatly in the coming decades, it is important to realize that much of what we would like to know about exoplanet properties cannot be obtained in this way. To take just one example, although it might in principle be possible to determine whether a rocky exoplanet is volcanically and tectonically active by observing trace gasses in its atmosphere using astronomical spectroscopy (e.g. Kaltenegger et al., 2010a), such observations will not be able to discriminate between different styles of volcanism (e.g. mantle plume vs. plate tectonics) which will be vital for understanding a planet's geological evolution (e.g. Stamenkovic and Seager 2016). Moreover, such observations would say nothing about the *history* of volcanic activity on the planet.

Multiple other examples readily spring to mind. For example, probing the detailed internal structure of a planet will require the application of geophysical techniques such as seismology and the measurement of local gravitational and magnetic fields; piecing together a planet's geological history will require imaging and spectral measurements of its surface with km-scale resolution, supplemented by mineralogical and geochemical measurements of the surface (and, ideally, sub-surface); dating key events in a planet's history will require precise measurements of radiogenic isotopes in a wide range of rock samples. Just as we have found in our own solar system, obtaining planetary data of this kind will require *in situ* measurements.

*Astrobiology*

The search for life on planets orbiting other stars is probably the most scientifically and publicly compelling aspect of exoplanet research. In this context, it is important to realize that, long before interstellar space travel becomes possible, advances in astronomical techniques will likely enable the detection of molecular biosignatures in the atmospheres and/or on the surfaces of nearby exoplanets (e.g. Seager 2014; see also other chapters in this Handbook). Indeed, one could argue that such a detection might provide the strongest of all motivations for the development of interstellar probes, to confirm the interpretation and to discover more about the alien biosphere. Note, however, that the absence of a detectable biosignature does not necessarily mean that life is absent (e.g. Cockell 2014). For example, an extraterrestrial version of the proposed *Darwin* space interferometer (Cockell et al., 2009) may not have found any evidence for life on Earth prior to the buildup of oxygen in the atmosphere about 2.3 billion years ago, yet life was certainly present much earlier (e.g. Knoll, 2004).

We can be reasonably confident that in the coming decades astronomical observations will be sufficient to establish a hierarchy of astrobiological priorities among planets orbiting the nearest stars: (i) planets where plausible atmospheric biosignatures are detected (e.g., Cockell et al. 2009; Kaltenegger et al. 2010b, Seager 2014, Hegde et al. 2015, and other chapters in this Handbook); (ii) planets that appear habitable (e.g. for which there is spectral evidence for water and carbon dioxide, but no explicit evidence of life being present); and (iii) planets which appear to have uninhabitable surfaces (either because of atmospheric compositions deemed non-conducive to life or because they lack a detectable atmosphere), but which might nevertheless support a subsurface biosphere. Thus, when planning an interstellar mission with astrobiology in mind, we are likely to have a prioritized list of target systems prepared in advance.

However, it is important to realize that even in the highest priority cases (i.e. exoplanets where *bona fide* spectral biosignatures have been detected), it may be difficult or impossible to prove that life is present. This difficulty has been well articulated by Cockell (2014) who noted that:

> "the lack of knowledge about an exoplanet, including plate tectonics, hydrosphere-geosphere interactions, crustal geochemical cycling and gaseous sources and sinks, makes it impossible to distinguish a putative biotic contribution to the [atmospheric] mixing ratios, regardless of the resolving power of the telescope …. The lack of knowledge about an exoplanet cannot necessarily be compensated for by improving the quality of the spectrum obtained."

Note how astrobiological considerations are here tied into the need for a comprehensive *geological* understanding of the planet under consideration which, as we noted above, will itself probably require *in situ* investigation.

Moreover, even if it does prove possible to demonstrate the presence of life on an exoplanet by means of telescopic observations from the solar system, gaining knowledge of the underlying biochemistry, cellular structure, ecological diversity, and evolutionary history of such a biosphere surely will not be obtained by such techniques alone. Only *in situ* investigations with sophisticated scientific instruments will enable biologists to study an alien biosphere in any detail. Furthermore, understanding the evolutionary history of such a biosphere will additionally require the tools and techniques of palaeontology, which will also depend on physical access to the environment.

## Implications for interstellar mission architecture

As noted above, long before interstellar spaceflight becomes feasible we are likely to know the basic structure of nearby exoplanetary systems and to have produced a prioritized list of systems suitable for *in situ* investigation. We also need to be aware that the sophistication of solar system-based astronomical instruments and techniques will increase considerably over the coming decades (e.g. Schneider et al. 2010; see also other chapters in this Handbook). Given the likely cost and complexity of building interstellar space probes, it will therefore be necessary to focus their capabilities on obtaining scientific information that cannot plausibly be obtained by utilizing remote-sensing techniques across interstellar distances. These considerations have implications for the architecture of an interstellar mission designed with planetary science and astrobiology in mind.

For solar system missions, there is a hierarchy of architectural options in order of increasing complexity and energy requirements, but also in increasing scientific return: (i) flyby missions; (ii) orbital missions; (iii) landers (with or without rover-facilitated mobility); and (iv) sample return missions. The same general ordering will apply in the study of extrasolar planetary systems, although the relative jumps in difficulty between them are not the same.

An undecelerated flyby will be the easiest to implement, and for this reason has been adopted in several interstellar mission studies (see below). However, the exploration of the solar system shows that flybys are very limited in terms of the knowledge they can collect, and that such information can be misleading (as in the case of the early flybys of Mars which revealed a lunar-like landscape and gave little intimation of the geological diversity discovered by later missions). Moreover, the limitations of flybys in an interstellar mission will be exacerbated by the necessarily very high speeds involved. Much more scientific information would be obtained if an interstellar vehicle could be decelerated from its cruise velocity to rest in the target system – the benefits will be immediately obvious by comparing the results of the initial flyby reconnaissance of Mars with those obtained by later orbital missions.

In the solar system, even more detailed information has resulted from the handful of soft landers and rovers that have successfully reached planetary surfaces. Although in terms of solar system exploration there is a big jump in energy requirements between orbital missions and soft landers, this would not be a major consideration in terms of an interstellar mission: the energy differential between orbital insertion and a soft landing is trivial in comparison to that of decelerating a spacecraft from a significant fraction of the speed of light. As for solar system missions, landers would permit a range of geochemical, geophysical and astrobiological investigations that will not be possible from an orbiting spacecraft. Thus, despite the added complexity, the potential scientific benefits are such that the designers of any interstellar mission capable of decelerating at its destination should consider including sub-probes that are capable of landing on the surfaces of suitable planets. This would be in addition to providing planetary orbiters (which will in any case be needed as communication relays if landers are deployed).

The most ambitious solar system missions involve sample return, which allows detailed investigation of planetary materials in terrestrial laboratories. However, for any reasonable extrapolation of foreseeable technology, this will not be possible from an extrasolar planetary system on any reasonable timescale. For this reason, it would be desirable for interstellar space probes to carry equipment able to make automatously the kinds of analytical measurements usually made in terrestrial laboratories.

Finally, it will be necessary for the results of scientific measurements to be transmitted back to the solar system. This is not the place to do a full trade study of an interstellar communications system, and the interested reader is referred to Lawton and Wright (1978), Lesh et al (1996) and Milne et al (2016). However, given the distances involved, ensuring a data-rate that would do justice to the richness of the proposed scientific investigations is likely to require transmitter powers in the MW range and, at least for radio communications, transmitting apertures tens to hundreds of metres in diameter. Laser-based communications systems may be preferable (e.g. Lesh et al. 1996), but any such system will, of necessity, still be quite massive and power-consuming.

These considerations have significant implications for the size and mass of proposed interstellar spacecraft. Maximizing scientific return will require a vehicle able to accelerate a capable scientific payload to a significant fraction of the speed of light, decelerate it again at the target system, and carry a communications system able to return the data. Moreover, sub-probes will be required that can land on planetary surfaces to conduct *in situ* investigations. Although considerable advances in automating and miniaturizing scientific instruments can be expected well before interstellar spaceflight becomes feasible, a meaningful scientific payload, able to make a comprehensive set of measurements of an exoplanetary system, will surely have a mass of at least several tonnes and possibly very much more (Crawford, 2016a). Moreover, the total mass that must be accelerated (and at least in part decelerated) will be substantially more than the scientific payload alone, owing to the mass of the propulsion system, structural supports, power supplies, communications equipment, protection against interstellar dust and, for most concepts, fuel with which to decelerate.

## A brief review if interstellar propulsion concepts

There is already a substantial literature devoted to the technical requirements of rapid interstellar spaceflight (where I take 'rapid' to imply velocities of the order of ten percent of the speed of light ($0.1c$), thereby permitting travel times to the nearest stars of several decades). Lower velocity options can also be envisaged, but these appear less relevant to the requirements of scientific exploration. Probably the first serious discussion of the topic was the ground-breaking paper on "Interstellar Flight" by Shepherd (1952), with other early contributions including those of Bussard (1960), Forward (1962), Strong (1965), Marx (1965), and Dyson (1968). Detailed reviews of the subsequent literature have been given by, among others, Mallove and Matloff (1989), Crawford (1990), Mauldin (1992), Frisbee (2009), Matloff (2010), Long (2012), and Lubin (2016).

Several proposed methods for achieving rapid interstellar spaceflight are discussed in the following sections. Readers interested in additional details are referred to the above-mentioned review articles, and to the more specialist papers cited under each topic below.

*Rockets*

Most space exploration to-date has been achieved using rockets, which carry their fuel with them. This leads to simplicity of design and operation, but has the fundamental limitation that energy must be expended to accelerate that portion of the fuel which has not yet been consumed. As a consequence, the ratio of the initial mass of the rocket to its final mass, the mass-ratio ($R$), rises exponentially with the velocity gained during its flight ($\Delta v$) in accordance with the rocket equation:

$$R = \frac{M_{veh} + M_{fuel}}{M_{veh}} = e^{\Delta v/v_e} \qquad (1)$$

where $M_{veh}$ is the mass of the vehicle without its fuel (i.e. the 'dry' mass; including the payload, engine, fuel tanks, and supporting structure), $M_{fuel}$ is the mass of the fuel, and $v_e$ is the rocket's exhaust velocity. Because $\Delta v$ must be very high for an interstellar rocket (~0.1c), realistic mass-ratios necessitate comparably high exhaust velocities. As the required exhaust velocities are much higher than can be obtained using chemical propellants, most proposed interstellar rocket concepts are based on nuclear power sources.

## Nuclear rockets

Although nuclear fission-powered rockets have long been studied in the context of interplanetary travel (e.g. Dewar 2007) they are probably not capable of achieving the velocities required for rapid interstellar flight (e.g. Freeland 2013), and most published concepts rely on nuclear fusion in one form or another.

The most detailed such concept available in the literature is still the Project *Daedalus* study conducted in the 1970s (Bond et al. 1978; Bond and Martin 1986). The aim was to design a vehicle capable of accelerating a 450 tonne payload to a cruise velocity of about 0.12$c$, which would result in a travel time to the nearest star of ~40 years. The resulting vehicle was a two-stage pulsed nuclear fusion rocket in which the fusion energy was magnetically contained within a reaction chamber and used to generate thrust. The study found that at total mass of 50,000 tonnes of nuclear fuel (consisting of pellets of deuterium and $^3$He), and a dry mass of about 2700 tonnes, was required to accelerate the second stage and payload to 0.12$c$ over a period of 3.8 years. The *Daedalus* vehicle was not designed to decelerate at its destination, which would greatly limit its scientific value, although the design could in principle be modified to use the second stage to decelerate at the expense of halving the maximum velocity and doubling the overall travel time.

Given advances made in miniaturization and nanotechnology since the *Daedalus* study, a payload mass of 450 tonnes now looks extravagant. However, note that (i) there is actually little advantage in making the payload less massive than the propulsion system so, unless the mass of the latter can be significantly reduced, it makes more sense to use advances in miniaturization to increase the capability of a payload (e.g. to include more probes and instruments) than to reduce its overall mass; and (ii) consideration of the requirements for a detailed study of an exoplanetary system, including multiple sub-probes to investigate different planets, indicates that payload masses of the order of 100 tonnes will probably be required in any case (Crawford 2016a).

It has become clear that some aspects of the *Daedalus* concept would probably be unworkable in practice (e.g. French 2013, Long 2016), and further studies are required. One such study is Project *Icarus* (Swinney et al. 2011), which aims to bring the *Daedalus* study up-to-date and which explicitly includes payload deceleration (albeit with the lower payload mass of 150 tonnes) in its design criteria. In addition to pulsed fusion concepts, the *Icarus* study is investigating propulsion concepts based on continuous fusion which may be more efficient (e.g. Freeland and Lamontagne 2015). Both approaches will continue to be aided by improved understanding of controlled nuclear fusion gained from experimental facilities designed to investigate the potential of fusion power on Earth, and continued refinement of the design concepts may be expected (Long 2016).

**Antimatter rockets**

Published studies of nuclear fusion rockets give some confidence that they will be able to deliver a scientific payload with a mass of order 100 tonnes to a nearby star system with a travel time of under 100 years. However, such studies also indicate that huge vehicles, requiring tens of thousands of tonnes of nuclear fuel, will be required. The only way to reduce the mass of an interstellar rocket able to achieve the same velocity is to increase the energy density of the fuel, and the only known potential fuel that exceeds nuclear fusion in this respect would be matter-antimatter annihilation (e.g. Forward 1982, Morgan 1982, Semyonov 2014). Such a capability is, of course, well beyond our immediate technological horizon but, should it ever prove practical, the very high energy density of antimatter would enable much less massive vehicles to achieve the same objective. It would especially enable the deceleration a scientific payload from a cruise velocity of order $0.1c$, something that is a key scientific priority but one that is much more challenging using fusion-based methods.

It is relatively easy to show that for reasonable assumptions (e.g. Crawford 1990) the most efficient operation of an antimatter rocket will involve the annihilation of ~10 kg of antimatter to heat ~4 tonnes of reaction mass for each tonne of dry vehicle mass accelerated to a velocity of $0.1c$. Thus, if we wish to accelerate a 100 tonne payload to $0.1c$ and bring it to rest again, using an antimatter rocket engine of comparable mass (which can only be an assumption at present), and a structural mass amounting to 20% of the dry mass, then we would need to annihilate about 18 tonnes of antimatter to heat about 7000 tonnes of reaction mass. Smaller payloads and/or lower speeds would require less antimatter, but would be scientifically less capable (and, again, note that there is little advantage in making the payload much less massive than the engine). Just to put this in perspective, the energy locked in 18 tonnes of antimatter ($1.6 \times 10^{21}$ J) is about 20 times the current annual global electricity generating capacity (Enerdata 2016). The only plausible source for this energy would be sunlight, necessitating the construction of large (~100 km in size) solar energy collectors.

Needless-to-say, there will be multiple difficulties in producing antimatter in the quantities required and in storing and handling it safely (e.g. Semyonov 2017). Both are well beyond present capabilities. Given the unavoidably high cost of producing antimatter, it would be desirable to minimize its use, and hybrid concepts might help achieve this. For example, nuclear fusion might be used for the acceleration phase of a mission, and antimatter for the deceleration phase, as this would significantly reduce the initial accelerated mass while also minimizing the use of antimatter. It may also prove possible to use small quantities of antimatter to catalyze nuclear fusion (Gaidos et al., 1999) thereby improving the efficiency of the latter.

## *Beamed power propulsion*

The energy requirements of a space vehicle might be reduced if it didn't have to carry its own source of kinetic energy as fuel, but instead had this energy beamed to it from an external source.

### Light sails

The most familiar beamed power propulsion concept is that of a light-sail, where the energy and momentum of photons would accelerate a scientific payload. Natural sunlight is not sufficient to achieve the velocities required for rapid interstellar spaceflight and, as first suggested by Forward (1962) and Marx (1966), an intense collimated beam of light, such as could be produced by powerful lasers, would be required. Detailed reviews of the concept have been given by Forward (1984) and Lubin (2016), and the concept is now being studied in detail by the recently initiated *Breakthrough Starshot* project (http://breakthroughinitiatives.org/Initiative/3).

Although light-sails avoid the problem of having to accelerate unused fuel, it is easy to show that very large amounts of energy will still be required if a scientifically useful payload is to be accelerated to quasi-relativistic velocities by this method. The acceleration, $a$, produced by a laser power, $P$, impinging on the surface of a mass, $M$, is given by:

$$a = \frac{(1+\eta)P}{Mc} \qquad (2)$$

where η is the reflectivity of the surface and $c$ is the speed of light. Here, $M$ is the total accelerated mass (including the mass of the sail, its supporting structure, and the payload). Although a large reflecting surface (i.e. a 'sail') is not explicitly required by the physics of laser propulsion, in practice one will be required for at least two reasons: (i) the incident energy must be distributed over a sufficient area such that the vehicle is not damaged by it; and (ii) the angular size of the reflecting surface must be sufficient to capture the incident beam as its distance from the energy source increases.

Clearly, there will be many technical challenges and trade-offs in minimizing the sail thickness, to reduce its mass, while at the same time ensuring maximum reflectivity. Moreover, although transmitted power could be reduced by reducing the acceleration, this would require a longer acceleration distance to achieve the same final velocity, and therefore larger (and hence more massive) sails and/or larger diameter transmitting optics (to keep the energy within the angle subtended by the sail). Some examples of these trade-offs have been discussed by Crawford (1990; see also Forward 1984 and Lubin 2016). For example, if we wish to accelerate a 450 tonne vehicle to $0.12c$ in four years (i.e. as envisaged by the *Daedalus* project), this implies an acceleration of 0.285 m s$^{-2}$ and Equation (2) yields a transmitter power of 20 TW. This is about seven times the current global rate of electricity production (Enerdata 2016) and, as in the case of antimatter fuel discussed above, the only plausible source for this energy would be sunlight.

An even greater difficulty lies with small angular size of the sail as seen from the solar system. If, following Crawford (2016a), we assume that the scientific payload itself will need to have a mass of ~100 tonnes, this leaves 350 tonnes for the sail and its supporting structure. If we further assume that the supporting structure has the same mass as the sail (i.e. 175 tonnes each), that the sail is totally reflective ($\eta=1$), has a density equal to that of aluminum (2700 kg m$^{-3}$) and a thickness of 1µm (possibly unrealistic, but more conservative than the 16 nm proposed by Forward (1984)), we arrive at a sail diameter of ~9 km. This would subtend an angle of ~$8\times10^{-7}$ arcsec when acceleration ends at a distance of 0.07 parsecs from the solar system. Diffraction-limited transmitting optics would have to have a diameter of ~300 km at optical wavelengths to keep the beam within the sail area at this distance. These challenges might be mitigated if the payload mass could be reduced (although it would then become scientifically less capable), or if the sails could be made much less massive (e.g. by using advanced materials; Matloff 2013), or if the higher accelerations, and thus shorter acceleration distances, could be employed (although this would imply higher power levels for a given mass).

The *Breakthrough Starshot* project (http://breakthroughinitiatives.org/Initiative/3) takes the latter approach to extremes, by proposing to accelerate nano-craft (with masses of a few grams and sails of a few meters in size) to ~$0.2c$ using 10-100 GW lasers and accelerations of ~$10^4$ ms$^{-2}$. There can be little doubt that, if pursued to the hardware stage, projects such as *Starshot* will develop capabilities that ultimately will be very enabling for interstellar spaceflight. Benefits will include the development of laser technology, new materials, miniaturization of instruments and, if implemented, *in situ* studies of the local ISM beyond the heliosphere. However, it must be doubted that the very small payload masses envisaged for the *Starshot* probes, and the fact that they will probably have no way to slow down at their destination (but see Heller and Hippke 2017), will permit detailed scientific studies of exoplanets. Indeed, probes of this kind may not reveal much more information than is likely to be observable using telescopes from the solar system in a similar timeframe.

The difficulty of decelerating a laser-pushed light sail at the end of its journey is a fundamental problem with the concept from the point of view of scientific exploration. Forward (1984) suggested a possible solution based on multiple reflections between nested light sails configured such that a solar system-based laser could be used for this purpose. However, not only is this quite complicated from an operational point of view, it would require transmitting optics in the solar system able to keep the beam focused onto the sails at the distance to the target star, in turn requiring transmitting optics thousands of km in size. An alternative means of deceleration might be to use electric and/or magnetic fields to transfer momentum to the ISM (e.g. Perakis and Hein 2016). However, this will require significant additional mass on the spacecraft (to generate the powerful electric and magnetic fields required) and, given the low density of the local ISM (Crawford 2011), would probably add many decades to the total mission duration owing to the very low deceleration rates. Moreover, the large superconducting coils ('magsails') required may be very vulnerable to damage by collisions with interstellar dust particles during this protracted deceleration phase.

It follows that, although laser-pushed light sails may be a practical means of launching small, lightweight probes on fly-by missions to the closest stars, they may not be the most practical means of delivering the kind of scientific payload required to perform detailed scientific investigations.

### Laser-powered rockets

The problems associated with constructing large sails (and their attendant mass) might be overcome of the energy required for acceleration was instead focused in a small volume of an interstellar vehicle to heat an inert reaction mass carried on board. Such a vehicle would then be a laser-powered rocket (Jackson and Whitmire 1978). The concept certainly has some advantages and is worthy of further study, although it still suffers from the problem of maintaining a highly-collimated energy beam over interstellar distances.

### Interstellar pellet stream

Singer (1980) suggested that a stream of electromagnetically accelerated pellets (each having a mass of a few grams and velocities of ~0.2c) might be used to transfer momentum to a larger interstellar vehicle. The main difficulties with this concept are again the need to maintain a highly-collimated beam of pellets, and the apparent inability of the method to allow deceleration at the destination. However, it is interesting to note that the speed and mass of Singer's pellets are similar to those envisaged by the laser-pushed *Starshot* probes, and the latter project might develop technologies relevant to the implementation of such an approach.

*Interstellar ramjets*

Some of the difficulties pertaining to both rockets and beamed power propulsion might be overcome if an interstellar vehicle could collect its fuel from the ISM while *en route*. This is the concept of an interstellar ramjet (Bussard 1960), in which interstellar protons would be collected by an electromagnetic scoop and used to power a nuclear fusion propulsion system. Following Bussard's analysis (see also the discussion by Crawford 1990), it can be shown that even a perfectly efficient interstellar ramjet would require scoop radii of hundreds or thousands of km to achieve accelerations (~0.1 ms$^{-2}$) sufficient to deliver a massive payload to a nearby star with a travel time of a few decades. There are also numerous other practical difficulties with the concept, including the difficulty of controlling a sustaining proton-proton fusion reaction in the plasma as it passes through the vehicle, and the probable extreme inefficiency of an electromagnetic scoop in collecting interstellar protons (e.g. Martin 1973, Heppenheimer 1978).

There is therefore a consensus that the basic ramjet concept as originally proposed is probably not realistic as a solution to the problem of rapid interstellar spaceflight. However, over the years a number of proposed improvements to the concept have been advanced, included suggestions for carrying nuclear catalysts to increase the efficiency of on-board fusion (Whitmire 1975); allowing the vehicle to collect the bulk of its reaction mass from the ISM but have it carry its own (nuclear or antimatter) energy sources (i.e. a 'ram-augmented interstellar rocket', Bond 1974); beaming energy to it from solar system-based lasers (i.e. a 'laser ramjet', Whitmire and Jackson 1977), and laying down pellets of fuel in advance long the trajectory to be followed by the vehicle so as to avoid reliance on the very tenuous local ISM (i.e. a 'ramjet runway'; Whitmire and Jackson 1977, Matloff 1979). It remains to be seen if any of these suggestions will prove workable, although it is interesting to note that the small, laser accelerated, payloads proposed for *Breakthrough Starshot* might lend themselves to laying down a trail of fuel packages as proposed by the 'ramjet runway' concept and this may be worthy of further study.

*More exotic suggestions*

Given that we are discussing developments that may still be a century or more in the future, we should be open to the possibility that unforeseen scientific and technological developments may render rapid interstellar spaceflight easier than we currently suppose. Clearly, we cannot sensibly discuss unforeseen developments, or those that contravene the currently understood laws of physics, but there are several exotic suggestions that may merit further research despite being well beyond the current technological horizon. These include the use of artificially generated atomic-sized black holes as ultra-compact energy sources (Crane and Westmoreland 2009), extracting energy from the quantum vacuum (Froning 1986), and developing self-reproducing nano-machines that would require very low masses to be sent to a target star system yet be capable of assembling the necessary scientific infrastructure on arrival (e.g. Tough 1998). Some of these more speculative ideas, and others, are reviewed by Millis and Davis (2009).

# Potential show-stoppers

It will be obvious from this discussion that achieving rapid interstellar spaceflight with payloads sufficiently complex to undertake a detailed scientific investigation of a nearby exoplanetary system will be an enormous challenge on multiple levels. We address three of the more serious potential problems here.

*Interstellar dust*

The ubiquitous presence of interstellar dust is often cited as the principal obstacle to rapid interstellar spaceflight (e.g. Schneider et al. 2010), and it is certainly true that impacts with interstellar grains will be potentially damaging. This was considered in detail in the context of the *Daedalus* study by Martin (1978), and Crawford (2011) extended this discussion in the light of more recent knowledge of the local interstellar dust density.

Martin (1978) adopted a beryllium shield for the *Daedalus* study owing to its low density and relatively high specific heat capacity, although further research could presumably identify better materials. Following Martin's analysis, but adopting a local interstellar dust density of $6.2\times10^{-24}$ kg m$^{-3}$ (Landgraf et al. 2000), we find that erosion by interstellar dust at a velocity of $0.1c$ would be expected to remove ~5 kg m$^{-2}$ of shielding material over a 1.8-parsec flight. The need to provide such shielding will certainly add to the mass of an interstellar probe, but should not present an insurmountable challenge. Clearly, the total mass of such a shield could be reduced by minimising the geometrical cross-section of the vehicle perpendicular to its direction of travel.

It is true that the upper boundary to the size distribution of interstellar dust particles in the solar neighbourhood is not well constrained (Landgraf et al 2000, Crawford 2011). If the local ISM contains a population of, as yet undiscovered, large grains (say tens to hundreds of µm in size, or even larger), then the individual kinetic energies of such particles would require special mitigation measures. For example, the kinetic energy of a 100 µm-diameter grain of silicate density moving at $0.1c$ (~$6\times10^5$ J) is equivalent to that of a 1 kg mass travelling at ~1 kms$^{-1}$. Clearly, more work needs to be done to determine the upper limit to the size distribution of interstellar dust grains in the local ISM. Future space missions beyond the heliosphere (possibly including precursor low-mass interstellar probes such as envisaged by *Breakthrough Starshot*; http://breakthroughinitiatives.org/Initiative/3), will be important in gathering this information before rapid interstellar spaceflight is attempted with significant scientific payloads.

Fortunately, even if a population of large interstellar grains is discovered in the local ISM, it is still possible to envisage appropriate countermeasures. For example, one could imagine using on-board sensing instruments (e.g. radar or lidar) to detect large incoming grains and to employ active (e.g., laser) or passive (e.g., electromagnetic) means with which to destroy or deflect them. However, probably the simplest solution, suggested by Bond (1978) for the *Daedalus* study, would be for the spacecraft to be preceded by a fine cloud of small dust particles (ejected from the vehicle and thus traveling at the same velocity but a small distance ahead), such that any incoming large grains would be destroyed by collisions within this artificial dust cloud before they have a chance to reach the main vehicle.

Thus, although the presence of interstellar dust certainly complicates proposals for rapid interstellar spaceflight it need not be a show-stopper. Schneider et al. (2010) are correct to point out that the mitigation strategies identified here would only apply to quite substantial vehicles – the only realistic mitigation strategy for low-mass probes, such as envisaged by *Breakthrough Starshot*, would be to minimize their cross-sectional area and to launch a large number in the hope that some will avoid collisions with interstellar dust. However, as we have argued above, *any* interstellar vehicle capable of delivering a scientifically useful payload to another planetary system is probably going to have to be quite massive anyway, and so is likely to have a mass budget available for a dust protection system.

*Sources of energy*

The main technical obstacle to rapid interstellar spaceflight is the very large amount of energy required to accelerate a useful scientific payload to a significant fraction of the speed of light. This will be placed in perspective if we consider that the kinetic energy of 100 metric tonnes moving at $0.1c$ is $4.5 \times 10^{19}$ J, or about half of current global annual electrical generating capacity (Enerdata 2016). When one considers the practical difficulties in sourcing this amount of energy, transferring it to a space vehicle, and the inherent inefficiencies of most propulsion concepts, the practical challenges will be all-too-apparent.

If this energy is to come from naturally occurring nuclear isotopes for use in nuclear fusion rockets then, as we have seen, many thousands of tonnes of these isotopes will need to be mined, refined and loaded onto a space vehicle. If the energy is to be provided in the form of antimatter then this will have to be manufactured using some other source of energy, and it will take at least as much (and probably many orders of magnitude more) energy to manufacture as will be gained from consuming it. Given the quantities of energy required this could only plausibly be obtained by collecting and processing solar energy. The same will be true of energy beamed to a laser-pushed light-sail or laser-powered interstellar rocket. In either case, space-based solar arrays hundreds of km on a side would be required, necessitating a significant space-based industrial capacity.

*Affordability*

Although very low-mass laser-pushed interstellar space probes might conceivably be launched directly from Earth in the coming decades with budgets comparable to other large-scale space projects (http://breakthroughinitiatives.org/Initiative/3), the scientific capabilities of such small payloads will surely be very limited. The much greater task of transporting a scientifically useful payload to nearby star is likely to require vehicles of such a size, with such highly energetic (and thus potentially dangerous) propulsion systems, that their construction and launch will surely have to take place in space. Moreover, as we have seen, obtaining the raw energy needed to accelerate such payloads to a significant fraction of the speed of light will also require significant space-based industrial capabilities.

It seems most likely that an interstellar spacefaring capability will only become possible in the context of a well-developed space economy with access to the material and energy resources of our own solar system (e.g. Hartmann 1985, Lewis et al 1993, Metzger et al 2013). Developing such a space economy will doubtless take many decades (and perhaps centuries), and is unlikely to be driven by long-term considerations of interstellar exploration (although the discovery of *bona fide* biosignatures in the atmosphere of a nearby exoplanet may give a boost to such activities). However, it should be noted that developing the kind industrial infrastructure in space that would ultimately permit the construction of interstellar spacecraft would enable many other scientific benefits in the meantime, including, in the present context, the construction of large space telescopes for the study of exoplanets (Crawford 2016b). Indeed, it is precisely because the economic development of space will render large space telescopes affordable well before rapid interstellar space probes become feasible that, if they are to be built at all, the latter must be able to perform the kind of *in situ* measurements that cannot be made by the former.

## Conclusions

Exoplanet research would be a major beneficiary of developing a rapid interstellar spaceflight capability because it would enable the investigation of other planetary systems with the same kinds of *in situ* techniques currently applied to the study of planets in our own solar system. Much of the information that might be gained by *in situ* investigation is unlikely ever to be obtained using astronomical remote sensing techniques, no matter how large or sophisticated future astronomical instruments may become. In the field of astrobiology, direct investigation using interstellar probes may be the only way to follow-up detections of putative biosignatures made in the atmospheres of Earth-like planets orbiting nearby stars.

It is clear that rapid interstellar spaceflight (here defined as achieving velocities ∼$0.1c$, and thereby enabling travel times to the nearest stars of several decades) will be a considerable technological and economic undertaking. The magnitude of the difficulties should not be underestimated, but neither should they be exaggerated. There is a large body of technical literature, partly reviewed above, demonstrating that rapid interstellar space travel is not physically impossible and is therefore a legitimate aspiration for the more distant future.

Although very low-mass laser-pushed interstellar space probes may become feasible within the next few decades, the scientific capabilities of such probes will be very limited. Indeed, it is not clear that the capabilities of such probes, which will have very limited communications capabilities and probably no means of decelerating at their destinations (but note Heller and Hippke 2017), will exceed what is likely to be achievable by astronomical techniques from the solar system in a comparable timeframe. That said, technologies developed in support of this approach, especially with regard to high-power lasers, focusing optics, low-power communications, and miniaturized instruments, and the possibility of such probes making direct measurements of the local ISM, will help pave the way for more ambitious and scientifically capable approaches. For this reason, initiatives such as *Breakthrough Starshot* (http://breakthroughinitiatives.org/Initiative/3) are greatly to be welcomed.

That said, barring spectacular (but perhaps not entirely unforeseeable) developments in nanotechnology, it seems that scientifically meaningful investigations of exoplanetary systems will require accelerating payloads with masses of many tonnes to a significant fraction of the speed of light, and decelerating them again at their destination. Whether this is done with nuclear fusion (or antimatter) rockets, laser-pushed light sails, interstellar ramjets, or some combination of these approaches, it is unlikely that such a capability will be possible until a significant industrial and engineering infrastructure has first been developed in the solar system. A space economy, based on utilizing the energy and raw materials of the solar system, may therefore be a prerequisite for rapid interstellar spaceflight. Such an economy may take several centuries to develop, and is unlikely to be driven by scientific considerations alone, but it would nevertheless enable many scientific opportunities in the solar system while also laying the foundations for an interstellar spaceflight capability from which exoplanet science will ultimately benefit.

## Acknowledgements

I would like to thank Jean Schneider for his invitation to write this review, and also to thank him for his work in creating and maintaining the invaluable Extrasolar Planets Encyclopaedia (http://exoplanet.eu/). I thank Bob Parkinson for first drawing my attention to *Project Icarus* some years ago, and Stephen Baxter, Kelvin Long, Rob Swinney, and other colleagues on that project, for encouraging me to think more deeply about the challenges and opportunities of interstellar spaceflight.